# Residual Metallic Contamination of Transferred Chemical Vapor Deposited Graphene

*Grzegorz Lupina,‡\* Julia Kitzmann,‡ Ioan Costina,‡ Mindaugas Lukosius,‡ Christian Wenger,‡ Andre Wolff,‡ Sam Vaziri,† Mikael Ostling,† Iwona Pasternak,§ Aleksandra Krajewska,§ Wlodek Strupinski,§ Satender Kataria,⊥ Amit Gahoi,⊥ Max C. Lemme⊥, Guenther Ruhl┬, Guenther Zoth,┬ Oliver Luxenhofer,∥ Wolfgang Mehr‡*

‡IHP, Im Technologiepark 25, 15236 Frankfurt (Oder), Germany

†KTH Royal Institute of Technology, School of ICT, Isafjordsgatan 22, 16440 Kista, Sweden

§Institute of Electronic Materials Technology, Wolczynska 133, 01-919 Warsaw, Poland

⊥University of Siegen, Hölderlinstr. 3, 57076 Siegen, Germany

┬Infineon Technologies AG, Regensburg 93049, Germany

∥Infineon Technologies Dresden GmbH, Dresden 01099, Germany

\*Correspondence to: Grzegorz Lupina, lupina@ihp-microelectronics.com

# ABSTRACT

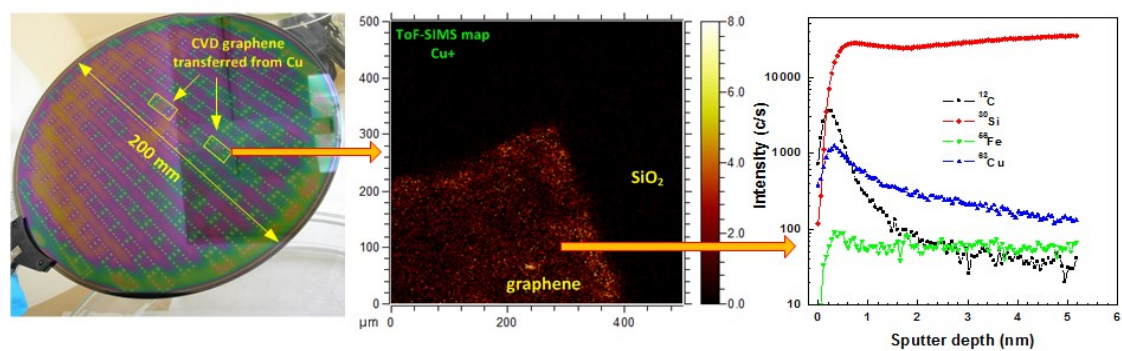

Integration of graphene with Si microelectronics is very appealing by offering potentially a broad range of new functionalities. New materials to be integrated with Si platform must conform to stringent purity standards. Here, we investigate graphene layers grown on copper foils by chemical vapor deposition and transferred to silicon wafers by wet etch and electrochemical delamination methods with respect to residual sub-monolayer metallic contaminations. Regardless of the transfer method and associated cleaning scheme, time-of-flight secondary ion mass spectrometry and total reflection x-ray fluorescence measurements indicate that the graphene sheets are contaminated with residual metals (copper, iron) with a concentration exceeding $10^{13}$ atoms/cm$^2$. These metal impurities appear to be partly mobile upon thermal treatment as shown by depth profiling and reduction of the minority charge carrier diffusion length in the silicon substrate. As residual metallic impurities can significantly alter electronic and electrochemical properties of graphene and can severely impede the process of integration with silicon microelectronics these results reveal that further progress in synthesis, handling, and cleaning of graphene is required on the way to its advanced electronic and optoelectronic applications.

Graphene has a high potential to provide performance boost for the next generation of high frequency electronic and photonic devices.[1-7] In view of these applications, chemical vapor deposition (CVD) on metal surfaces is currently one of the most relevant graphene synthesis techniques delivering large area and good quality material.[8] Practical use of transferred CVD graphene in electronic and photonic devices will likely require a co-integration of the new material with the existing semiconductor device manufacturing platforms. For example, CVD graphene will have to comply with very stringent purity standards. A large research effort has been dedicated so far to study residual polymer impurities resulting from graphene transfer.[8] Significantly less attention has been paid to potential sub-monolayer metallic contamination of graphene associated with the growth on and transfer from metal catalysts as for example Cu or Ni. Since trace impurities in silicon can result in detrimental effects on the performance of electronic devices, detection and control of metal contaminants in Si integrated circuit manufacturing is of critical importance to achieve high product yield. The effects of metal contamination (*e.g.* Cu, Ni, Fe) include junction leakage current increase, lifetime and dielectric strength degradation.[9] Even at very low concentrations ($10^{10} - 10^{11}$ atoms/cm$^2$) trace metals pose a serious threat to Si devices.[10] Since CVD graphene is usually synthesized on metallic surfaces, the growth and transfer processes can potentially cause residual contamination of graphene sheets. Although graphene is reported to be an effective barrier against Cu diffusion,[11] residual Cu atoms from contaminated graphene can potentially out-diffuse towards the substrate in the course of further device processing and result in degradation of device parts located beneath graphene. This can be expected based on the ability of Cu atoms (ions) to diffuse (drift) through dielectrics under thermal or electrical stress.[12,13] Residual metals released during device processing can also cause cross-contamination of sensitive manufacturing tools. Finally, it has been demonstrated that residual metallic impurities can significantly alter electronic and electrochemical properties of graphene.[14-16] It has been also shown that even if nuclear purity graphite is used as the source material for graphene synthesis, the latter can be contaminated with impurities originating from chemical reagents used for processing.[17]

The presence of metallic impurities on graphene transferred from Cu substrates was recently confirmed using various techniques such as inductively coupled plasma mass spectrometry, x-ray energy-dispersive spectroscopy, energy electron loss spectroscopy, and x-ray photoelectron spectroscopy (XPS) in several publications.[14,18,19] It has been shown that a wet Cu etching combined with a modified standard clean 2 (SC-2) used in Si device manufacturing,[20] reduces the concentration of residual metals below the detection limit of XPS tools[19] being about 0.1 at%.[21] However, to verify if stringent purity standards of Si integrated circuit (IC) manufacturing lines are met, investigations with more sensitive techniques are required.[22]

RESULTS AND DISCUSSION

The facts listed above motivate our study of residual transfer-related metallic contamination (Cu, Fe, etc) of large area CVD graphene. We investigate different transfer methods involving different polymer support films combined with various strategies of detaching graphene from the metal catalyst substrate (using different Cu etchants and electrochemical delamination). We use time-of-flight secondary ion mass spectrometry (ToF-SIMS) and total reflection x-ray fluorescence (TXRF) to obtain elemental fingerprints of residual contamination with a sensitivity better than $10^9$ atoms/cm$^2$. ToF-SIMS offers the capability of high-resolution elemental or molecular fragment mapping, however, due to the strongly varying matrix-dependent ionization cross sections it is very difficult to quantify the measured elemental concentrations. TXRF, on the other hand, is easily quantifiable but does not yield spatial resolution. Thus we calibrated ToF-SIMS using reference graphene samples measured by TXRF before. In this way, we experimentally demonstrate that even extensive wet chemical cleaning procedures fail to remove residual Cu completely and that there is a trade-off between the purity of the fragile graphene layer and its structural integrity. Furthermore, our results indicate that Cu impurities transferred along with graphene onto Si wafers can negatively affect the minority carrier diffusion length in the Si substrate. Experiments presented here were performed on graphene samples grown and transferred in several different laboratories. These set was complemented with samples grown and transferred by commercial graphene material manufacturers and suppliers.

About 1x1 cm$^2$ pieces of CVD graphene on Cu foils (for details see methods) were transferred onto three kinds of substrates: 300 nm SiO$_2$/Si(100), p-Si(100) wafers covered with native SiO$_2$, or patterned p-Si(100) substrates with Si pillars embedded into SiO$_2$. Different polymers such as poly(methyl methacrylate) PMMA and polystyrene (PS) were used as support during the transfer process. There was no clear influence of the type of the polymer on the concentration of residual metallic impurities. Unless explicitly stated otherwise data reported here refer to PMMA-supported transfer. To detach the graphene layer from the Cu substrate electrochemical delamination (EC)[23] and wet etching[24,25] using ammonium persulfate (APS), FeCl$_3$, and H$_2$SO$_4$ solutions was performed (see Methods).

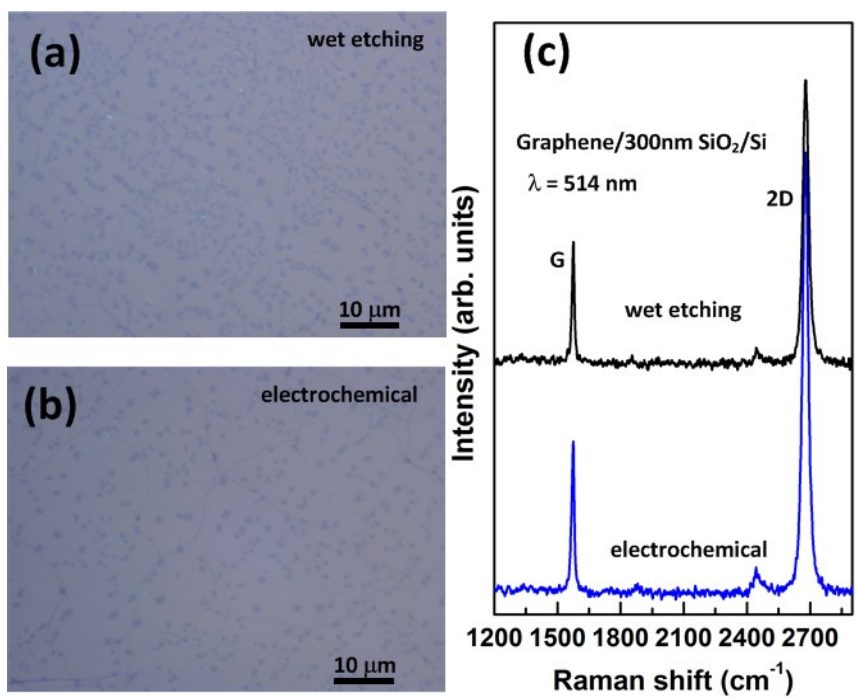

**Figure 1**. Comparison of graphene layers transferred using wet etching and electrochemical delamination. (a-b) Optical microscope images. Dark spots are multilayer graphene islands. (c) Corresponding Raman spectra.

The quality of the transfer process for each sample was controlled with optical microscopy (OM) and Raman spectroscopy. Figure 1 (a,b) shows OM images of graphene layers transferred onto 300nm $SiO_2$/Si using wet etching and electrochemical delamination. Figure 1 c shows the corresponding representative Raman spectra. In general, both transfer techniques result in good quality graphene with a low amount of cracks and holes and a low intensity Raman D band (see also Supporting Information Figure S1 and S2). For ToF-SIMS measurements, the 1x1 $cm^2$ graphene patches were inspected with OM and Raman spectroscopy and the areas with the best quality (*i.e.* low amount of holes and particles, low Raman D band) were selected for further investigation.

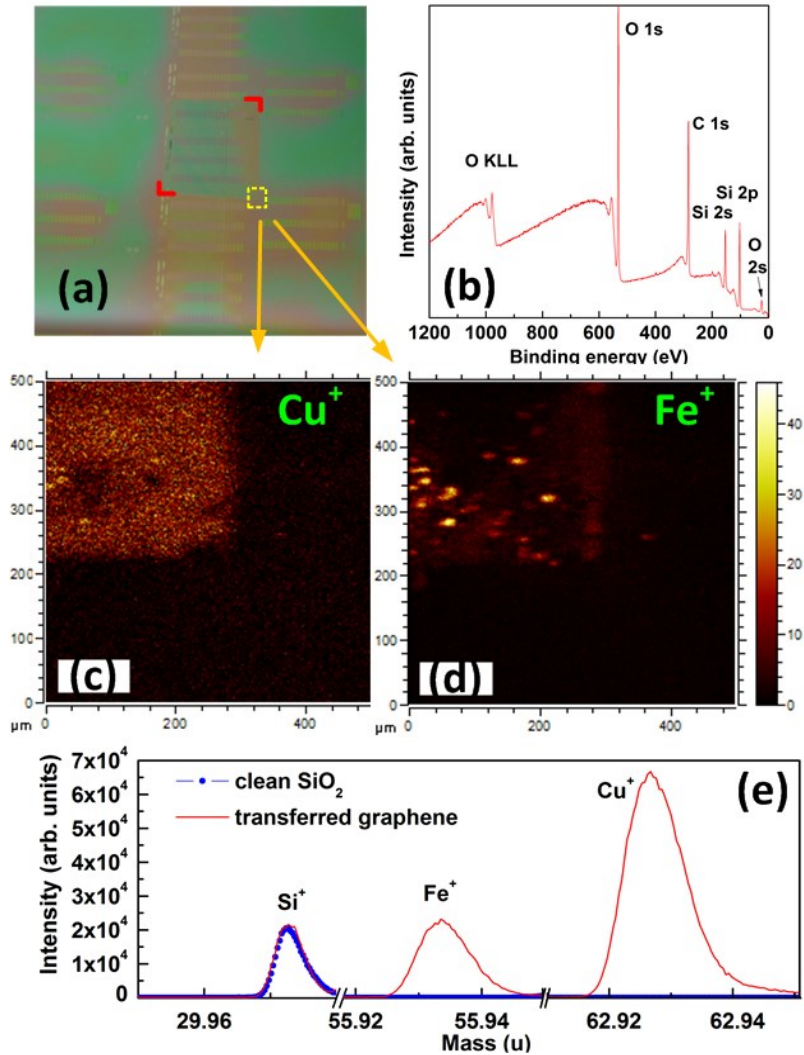

**Figure 2**. Residual metallic contmainations on CVD graphene. (a) a photograph of ~1x1cm$^2$ large graphene flake transferred to a patterned Si chip. (b) XPS overview scan on the area covered by graphene. (c-d) ToF SIMS $^{63}$Cu$^+$ and $^{56}$Fe$^+$ maps on the corner of the graphene layer. ToF SIMS mass spectra in selected regions acquired on transferred graphene and on a clean SiO$_2$ reference sample (e). Spectra are normalized to the intensity of the $^{30}$Si peak.

Figure 2 presents an illustrative example of ToF SIMS investigations performed on a graphene layer transferred onto patterned substrate (Figure 2a) using FeCl$_3$-based wet etching method. While XPS (Figure 2b) does not detect any metallic species on the surface, the ToF SIMS maps (Figure 2c-d) acquired in the bunched mode show clear evidence of Cu and Fe residuals on the areas covered with graphene (see Supporting Information for additional XPS results). Comparison of ToF SIMS mass spectra for a thermally grown 300nm SiO$_2$/Si substrate without and with graphene (Figure 2e) proves that the presence of residual metals is related to the graphene transfer process itself. In general, the regions close to the edge of the graphene flake appear to be more heavily contaminated showing relatively large agglomerations of metallic impurities. This can be caused by the mechanical deformation of the PMMA/graphene/Cu stack during cutting of the graphene/Cu stack into smaller pieces. As the edge regions were found to be non-representative of the sample, all further measurements were performed on the areas located at least 500μm from the graphene edge to enable reliable analysis and meaningful comparison between different samples.

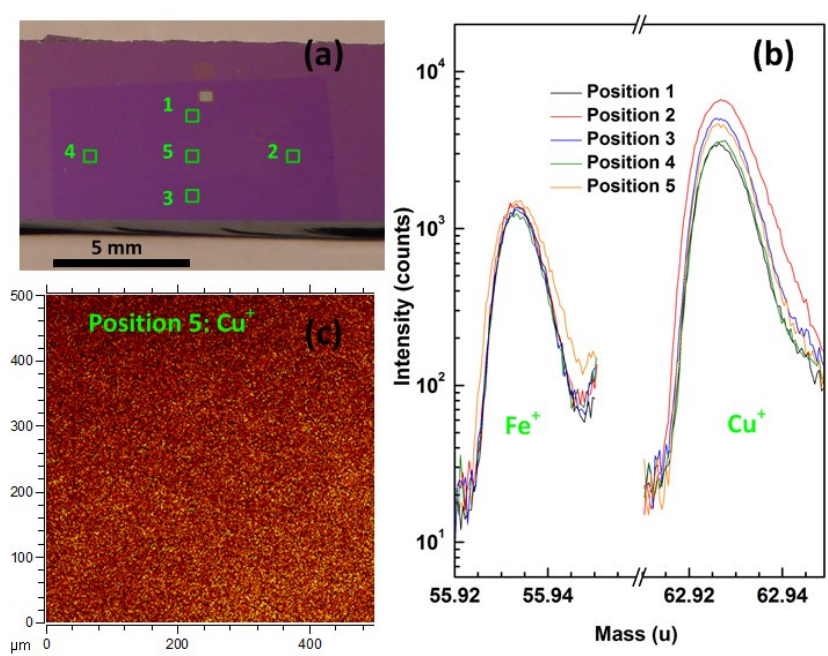

**Figure 3.** Distribution of metallic contaminants on the surface. (a) Optical microscope image of graphene layer transferred onto 300 nm $SiO_2$/Si substrate. (b) ToF SIMS mass spectra in $^{56}Fe^+$ and $^{63}Cu^+$ regions acquired at different points across the sample. Positions refer to the areas marked in panel (a). (c) 500x500 μm$^2$ ToF SIMS map of $Cu^+$ in the center of the sample. Measurements in panel b have been normalized to the $^{30}Si^+$ peak intensity.

Example of such measurement on 5 different spots across the sample is illustrated in Figure 3a. Mass spectra in the $^{56}Fe^+$ and $^{63}Cu^+$ regions (Figure 3b) indicate that the intensity of the $Cu^+$ and $Fe^+$ signals is relatively uniform. Also the individual ToF-SIMS maps acquired at different positions indicate, in contrast to the edge regions, a homogenous distribution of metallic contaminants within the mapping area of 500x500 μm$^2$ (Figure 3c). Larger Cu agglomerates were observed only occasionally.

Figure 4 summarizes the Cu surface concentration values measured with ToF SIMS calibrated to TXRF for graphene samples obtained by different detachment methods. We found that there is a broad distribution of Cu surface concentration values ranging from $10^{13}$ atoms/cm$^2$ to $10^{15}$ atoms/cm$^2$ for various graphene sources and transfer techniques. Furthermore, the amount of Cu residuals does not strongly depend on the type of the Cu etchant. Separate experiments confirmed that the chemicals used (VLSI grade) in various transfer processes were free of Cu traces (within detection limit of ToF SIMS). The lowest concentrations of residual Cu are found on samples etched in APS. However, this group of samples shows also the largest distribution of results. On samples etched in FeCl$_3$ apart from Cu a significant amount of Fe residuals was found. Although the latter can be quite effectively removed by the modified SC-2 clean,[19] the FeCl$_3$-based etchant does not present any advantage over APS and electrochemical delamination in terms of residual Cu. For this reason, only Fe-free detachment methods were used in further experiments. Interestingly, very similar amounts of Cu were found on graphene layers prepared by electrochemical delamination and wet etching. For some sorts of starting graphene material, electrochemical delamination produced heavily contaminated graphene samples (supporting information, Figure S4). This suggests that the delamination process should be adjusted to a given graphene sort to obtain a clean detachment. Subsequent treatment of delaminated graphene in for example HCl solutions usually reduced slightly the residual Cu amount to a level below $5\times10^{13}$ atoms/cm². However, samples with Cu impurity level lower than $10^{13}$ atoms/cm$^2$ could not be obtained with any of the mentioned methods. At the same time, measurements on the SiO$_2$ substrate in direct neighborhood of the flake (~0.5 - 1mm away from the graphene edge) indicated contamination of $10^{11}$-$10^{12}$ Cu atoms/cm$^2$.

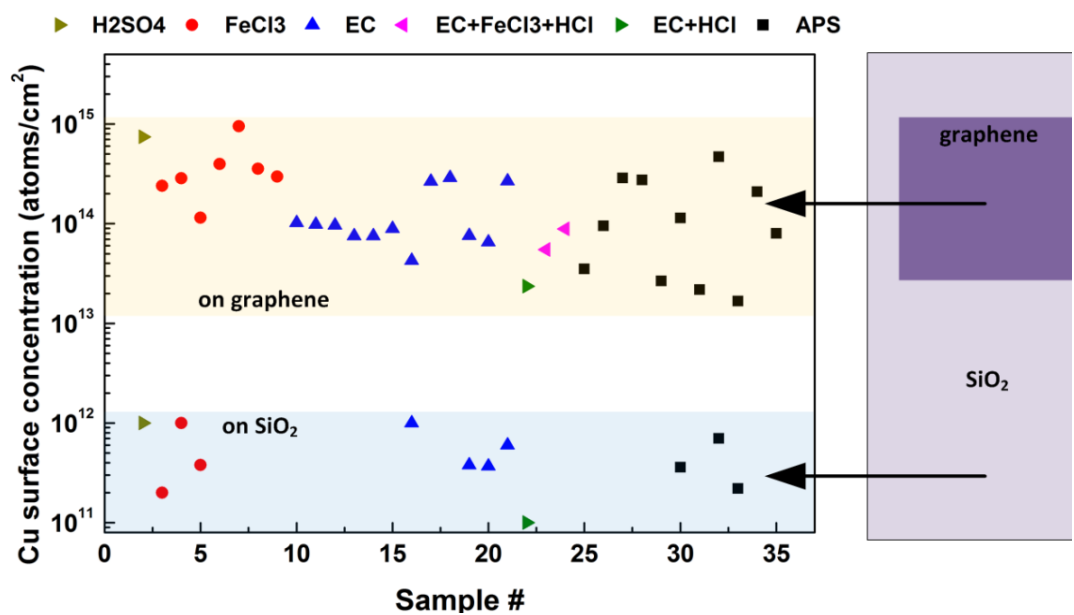

**Figure 4**. Comparison of surface concentration of Cu for different transfer methods. Measurements on graphene were performed in the center of the flake. Control measurements on the SiO$_2$ substrate were performed ~0.5 - 1mm away from the edge of the graphene flake.

Figure 5 presents results obtained for optimized transfer and cleaning protocols and thus constituting the cleanest graphene samples obtained in this work on 200 mm wafer substrates. Here, APS was used to etch Cu foil and the PMMA/graphene stack was rinsed several times in DI $H_2O$. Subsequently, samples were placed in an HCl-based cleaning solution to remove metallic residuals from graphene. Finally, graphene with PMMA was moved to a large volume container with DI $H_2O$ and transferred to the target wafer. A special care was taken to eliminate all metallic tools (tweezers, scissors *etc.*) from the transfer process. The concentration of Cu in the neighborhood of the graphene patches (indicated by arrow, S) is below the detection limit (BDL) indicating that the process does not contaminate the wafer with Cu beyond areas covered with graphene (for example by re-deposition during final rinsing step). At the same time, the concentration of Cu impurities on the areas covered with graphene is still exceeding $1 \times 10^{13}$ atoms/cm$^2$. Moreover, significant amounts of Fe residuals are found both on graphene and on the uncovered $SiO_2$ surface.

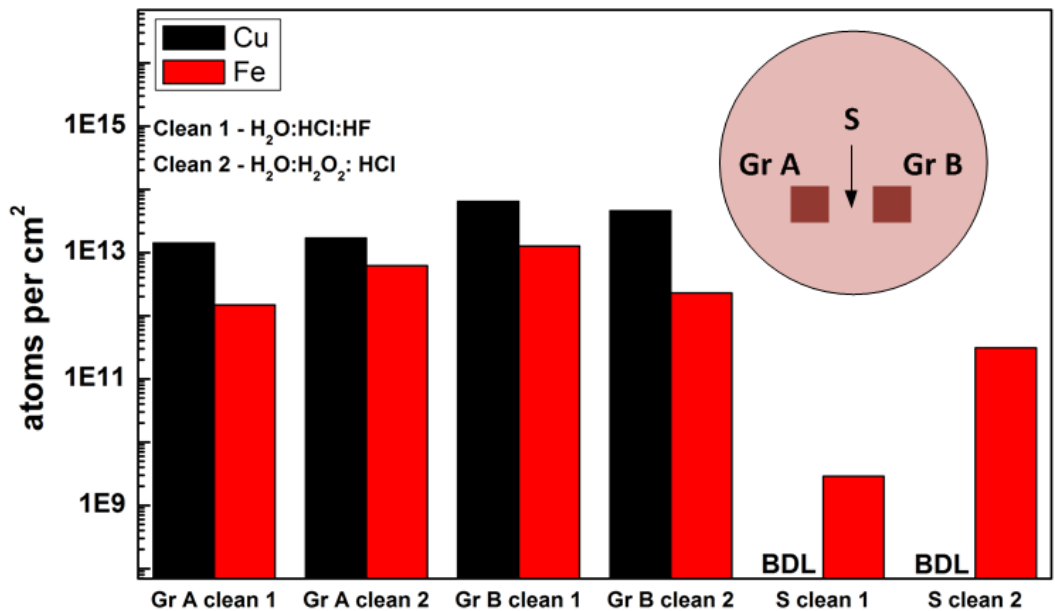

**Figure 5**. Surface concentration of Cu and Fe impurities measured by TXRF for two graphene samples from different sources (Gr A and Gr B) transferred onto 200 mm wafers using optimized transfer/cleaning protocols. BDL (below detection limit) indicates that no Cu impurity was detected in the neighourhood of the graphene patches (indicated by arrow, S).

Given this evidence it can be speculated that at least part of residual Cu may be "enclosed" into the graphene layer making it inaccessible for Cu etchants. Sublimation of Cu during graphene growth process and formation of graphene wrinkles during cooling[26,27] could for example result in Cu atoms being trapped in graphene pockets isolated from the Cu substrate. We also did not find a clear correlation between the Cu foil temperature during graphene growth and the amount of impurities after growth (supporting information, Figure S8). Similarly, we were unable to resolve any accumulations of Cu atoms which could be associated either with grain boundaries, wrinkles, graphene adlayers, or any other morphological features. An example of such attempt is illustrated in Figure 6 which shows a secondary ion (SI) image and a corresponding ToF SIMS $^{63}$Cu$^+$ map acquired in the burst alignment mode on the area of 20 x 20 μm$^2$. Bright dots visible in the SI image (Figure 6 a) correlate well with the multilayer islands visible in optical microscope images (compare Fig. 1). In the $^{63}$Cu$^+$ map (Figure 6 b) acquired on the same position these areas (examples marked with green circles) appear to have lower surface concentration of Cu than the monolayer graphene regions between the islands (red circles). This, however, may be also a consequence of the fact that in this mapping mode information is collected from the topmost surface layer only. As a result, Cu impurities under thicker graphene islands do not contribute to the acquired Cu distribution image as strongly as those located below monolayer graphene regions. Yet we found it difficult to resolve this with a sequence of mapping-sputtering steps (supporting information, Fig S5).

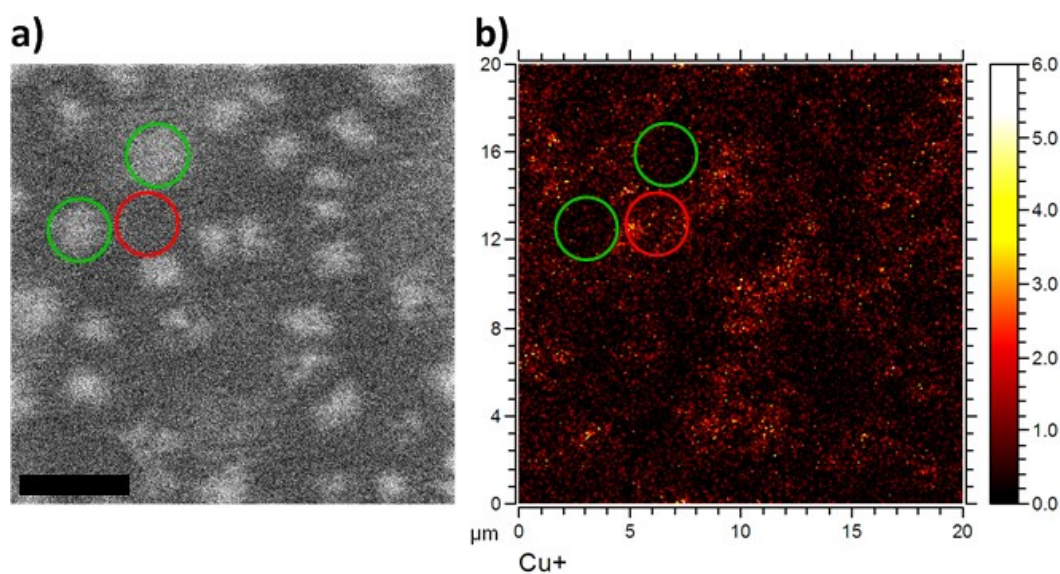

**Figure 6**. High lateral resolution ToF SIMS imaging on optimzed samples. (a) Secondary ion image showing multilayer graphene islands (brights spots) on monolayer graphene (dark background). The scale bar is 5 μm. (b) surface map of $^{63}$Cu$^+$ on the same region showing a correlation between the position of the islands and the surface concentration of Cu. Green and red circles mark the corresponding positions on both images.

In an attempt to further reduce the surface concentration of Cu, we investigated the effect of prolonged etching time in APS. According to these experiments, there is a significant difference in the amount of Cu present on the surface of samples etched for 8h and 72h, as shown in Figure 7a. Surface concentration of Cu decreases as the result of longer etching time by 50%. This apparent improvement in purity comes, however, at the expense of graphene layer integrity. As illustrated by optical microscope images in Figures 7 b and 7 c, prolonged contact with the Cu etchant results in the appearance of cracks visible as a network of bright lines in Figure 7c. Graphene seems to be washed away in these areas and the layer becomes discontinuous. This conclusion is in line with Raman map measurements performed on the sample etched for 72h (Figure 7d). The intensity of the 2D peak vanishes in some areas (black) reproducing a network of lines observed in optical images. Although it appears to be cleaner according to ToF-SIMS, graphene layer clearly becomes patchy after prolonged etching.

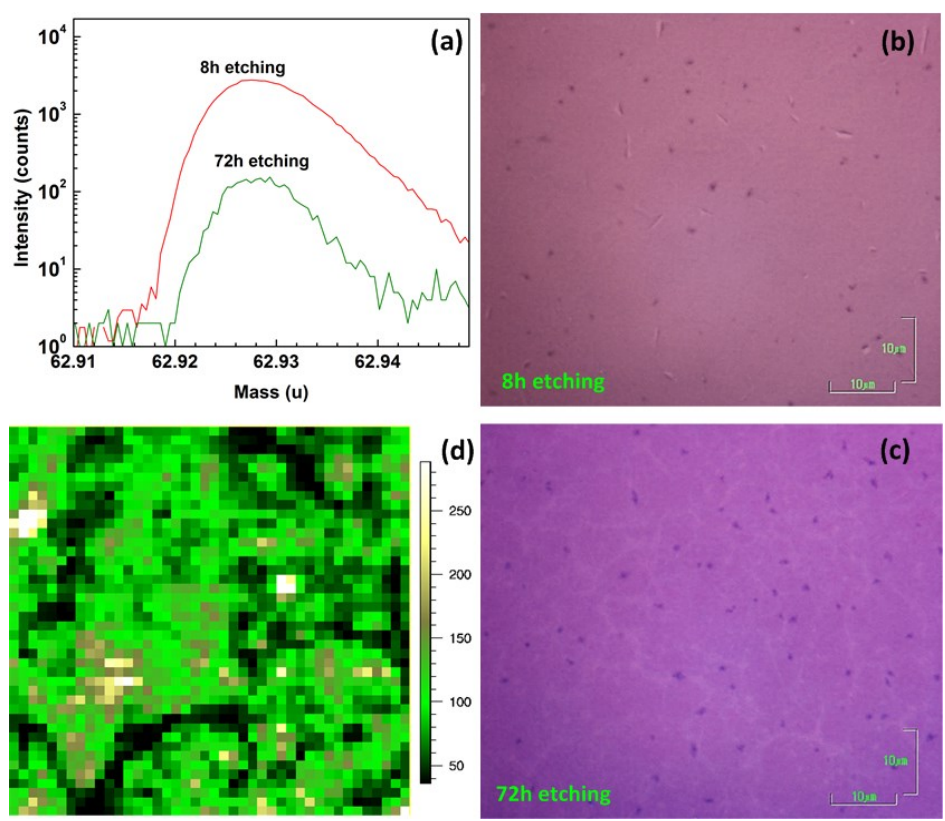

**Figure 7.** Investigation of prolonged wet etching time on the concentration of Cu residuals and the quality of graphene layers. (a) ToF SIMS mass spectra in the $^{63}$Cu region for samples with different etching time in APS solution. (b-c) Optical microscope images of graphene layers transferred to SiO$_2$ substrates after 8 and 72h etching. (d) Raman 2D peak intensity mapping on a sample etched for 72h. Mapped area is 20x20μm$^2$.

On the basis of the above observations, one can conclude that at least a part of the residual Cu atoms is trapped within the graphene layer and remains insensitive to etching and cleaning treatments. In this way, CVD graphene is transferred to the target substrate along with trace amounts of metallic contaminations. To asses if the residual metal is mobile and may out-diffuse during subsequent thermal treatment we performed several annealing experiments followed by ToF SIMS and minority charge carrier diffusion length measurements.

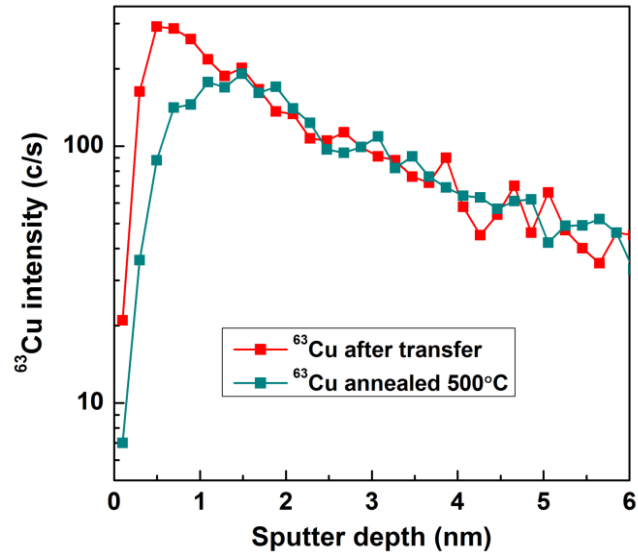

**Figure 8.** Influence of annealing on the amount of Cu residuals. Sample transferred onto native $SiO_2$/Si substrate was annealed in UHV at 500°C for 30min. Sputtering was performed with 0.5 keV Cs ions.

Figure 8 shows ToF SIMS Cu+ profiles from a graphene sample transferred onto native $SiO_2$/Si substrate and annealed subsequently at 500°C in UHV ($10^{-8}$ mbar) for 30min. Comparison of profiles taken before and immediately after annealing indicates that the thermal treatment resulted in a substantial reduction of the Cu concentration on the surface. This may imply a partial release of the Cu atoms from graphene and their diffusion into the underlying Si in line with the ability of Cu ions to penetrate thin $SiO_2$ layers.[28]

Minority charge carrier diffusion length (or the corresponding carrier lifetime) measurement is extremely sensitive to smallest amounts of impurities and hence an ultimate method for characterization of material quality and process control. It is widely used in silicon IC manufacturing for monitoring heavy metal contamination and key IC processing steps. The minority carrier lifetime is defined as the average time it takes an excess minority carrier to recombine. Low minority carrier lifetimes (low diffusion lengths) can be an indicative of metal contamination. In particular, Cu precipitates were reported to be extremely efficient minority carrier recombination sites.[29] To investigate the potential influence of the residual Cu on the minority carrier lifetime in the Si substrate two pieces of graphene from different suppliers were transferred onto p-type Si(100) wafer covered with native $SiO_2$ (Figure 9a). After transfer, the wafers with graphene were annealed at 600°C for 5 min in $N_2$. Subsequently, carrier diffusion length was measured point by point to create a map of the wafer shown in Figure 9 b.

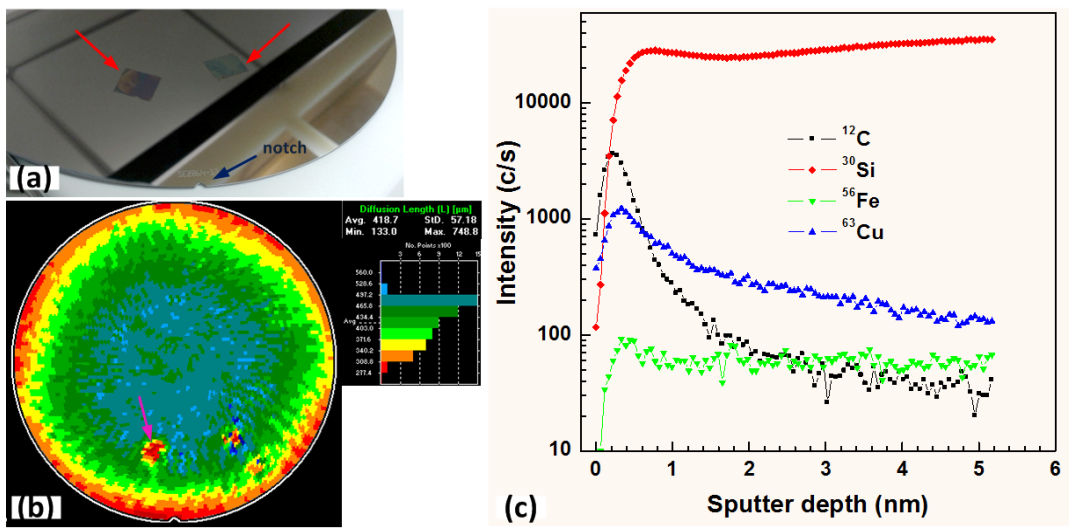

**Figure 9.** (a) Photograph of graphene layers transferred onto native $SiO_2$/Si(100) wafer. Photograph taken before PMMA removal, red arrows indicate the graphene locations. (b) Minority carrier diffusion length measurements on p-type Si wafer with graphene flakes after annealing at 600°C for 5min and optical activation of impurities.[30] (c) ToF-SIMS sputter profile on the graphene flake indicated by purple arrow in panel b. Sputter depth profiling was performed with 1keV $O_2$ ions.

The diffusion length map (Figure 9b) has been measured after optical activation of Cu and Fe impurities.[30] It clearly shows a significantly reduced diffusion length at exactly the two regions where the graphene flakes have been deposited (the strongly reduced diffusion length at the edge of the wafer is due to an unintentional contamination of the wafer edge and is not further considered). At these spots the diffusion length is about 300-350 µm compared to about 500 µm of a reference region at the wafer center. After storage of the sample at room temperature for several days a repeated diffusion length measurement showed a nearly complete recovery of the diffusion length at the two spots to values close to the values of the reference region.

Further measurements after optical or thermal treatment (200°C/5 min) resulted again in a significantly reduced diffusion length at the two regions. This reduction is again reversible as further measurements showed. From these observations the following conclusion can be drawn. Part of the diffusion length drop is due to Cu impurities (small precipitates) that usually form after optical activation of Cu-contaminated p-type Si and that reduce the diffusion length of minority carriers. This reduction is not reversible.[30] The major part of the diffusion length reduction stems from Fe impurities since we observe the typical recovery of the diffusion length during storage at room temperature and the reversible reduction after repeated optical or thermal activation of Fe impurities.[31] From the diffusion length change an Fe concentration in the range of $10^{10}$ atoms/cm$^{-3}$ can be estimated. We note that samples for this experiment were prepared by etching Cu in APS (FeCl$_3$-free transfer process) and that the origin of the Fe contamination is currently unknown. One of the possibilities is unintentional contamination during graphene growth process or handling as shown by our control measurements on as-shipped samples prior to transfer (supporting information, Figure S9).

ToF SIMS profiles shown in Figure 9c taken from one of the places covered by graphene (indicated by purple arrow in Figure 9b) appear to corroborate this claim. The $^{63}$Cu profile after annealing shows a high concentration in the upper few nm and approach the iron concentration in the bulk. In analogy to the case illustrated in Figure 8, the concentration of Cu at the surface decreased by about 40-50% as a result of the annealing treatment. Results of a reference experiment in which native SiO$_2$/Si surface was intentionally contaminated with Cu to a nominal level of about $5 \times 10^{14}$ atoms/cm$^2$ indicate that unbound Cu diffuses easily into Si and its concentration drops by two orders of magnitude after subjecting to a similar thermal budget (supporting information, Figure S6). The amount of Fe and Cu found in the graphene on the surface is more than enough to explain the diffusion length measurements in the bulk after anneal. So no clear quantitative conclusion can be drawn about the mobility of the contaminations. However, it can clearly be stated that Cu and Fe contaminations in graphene are in principle mobile to a significant amount. Considering the fact that already much lower Cu surface contaminations (~$10^{12}$ atoms/cm$^2$) can cause a far stronger degradation of minority carrier diffusion length in n-type Si than in p-type Si,[29] results presented here call for more attention and further studies in this direction.

CONCLUSIONS

In semiconductor device manufacturing platforms contamination control is absolutely essential, since even small amounts of impurities can result in altered device parameters, reliability and yield problems. This is reflected for example in the stringent specifications for high purity raw materials. Here, we investigated the purity of large area CVD graphene transferred from Cu to SiO$_2$/Si substrates with particular attention to a sub-monolayer Cu contamination. Our experiments show that regardless of the transfer method and subsequent cleaning trace amounts of metals (~$10^{13}$ – $10^{14}$ atoms/cm$^2$) are found on CVD graphene transferred to the target wafer. In the back-end-of-line (BEOL) integration of graphene devices[3], such contaminations may not play a significant role as most of the modern BEOL metallization layers are Cu-based. However, even such small amounts may be relevant when front-end-of-line (FEOL) integration approaches for electronic and photonic devices in Si IC fabrication lines are considered. In such a case, metallic impurities can lead to the contamination of Si devices and cross-contamination of fabrication tools. We find that a part of the residual Cu atoms can be released upon thermal treatment and out-diffuse affecting the minority carrier diffusion length in the Si substrate. According to our findings, the amount of impurities on graphene varies depending on the source of graphene and in consequence may be dependent on the CVD process used for graphene synthesis. These results call for more attention to the topic of sub-monolayer metallic contaminations of graphene and its influence on the performance of devices based on this new material. Clearly, further improvements in the transfer and cleaning technology are required to provide material of high quality and purity as demanded by microelectronic applications. This includes also the investigation of alternative metal catalyst-free paths to the fabrication of graphene directly on insulators and semiconductors.[32-34]

METHODS

**Sample preparation**

In the experiments described above we used either various sorts of commercially available graphene on Cu or graphene grown in our laboratories. In the latter case, 4 cm × 2 cm pieces of 25 µm thick Cu foil (AlfaAesar) was ultrasonically cleaned in acetone and was rinsed using IPA. The foil was loaded in NanoCVD chamber (Moorefield, UK). The chamber was purged five times using 200 SCCM of Ar gas and the chamber pressure was brought to less than 10 mTorr using a scroll pump. The Cu foil was then heated to 900 °C in Ar (190 SCCM) and H$_2$ (10 SCCM) atmosphere in 2 mins. Cu foil was kept under the same conditions for another 2 mins. The temperature was then increased to 950 °C in 60 s. Then the chamber pressure was increased to 10 Torr using Ar (80 SCCM) and H$_2$ (20 SCCM) mixture. At these conditions, Cu foil was annealed for 10 mins at 950 °C. Graphene growth was carried out at 950 °C for 30 mins using 5 SCCM of CH$_4$. After the growth, chamber was cooled down to room temperature in 90 mins under Ar atmosphere.

To transfer graphene, PMMA or PS solution was spin-coated on graphene/Cu stack. Graphene layer on the backside of the foil was removed by oxygen plasma etching. Ammonium persulfate (20-50mg/mL in water), iron(III)-chloride (80-120mg/mL in water), and 2:1:1 solution of H$_2$O:H$_2$SO$_4$:H$_2$O$_2$ were used to wet etch copper. Polymer/graphene stack was moved to distilled water several times to rinse the etchant residue. Electrochemical delamination was performed using NaOH or KCl as the electrolyte. Detached polymer/graphene stack was subsequently transferred to the target substrate, immersed in acetone bath to remove polymer support and finally the wafer with graphene was rinsed in isopropyl alcohol (IPA).

For a reference, we also measured CVD graphene grown and transferred by graphene material manufacturers and suppliers. These layers showed good crystalline quality and comparable level of metallic impurities as reported here.

**Characterization**

Raman spectra were acquired with a Renishaw InVia micro-Raman spectrometer equipped with a 514 nm (2.41 eV) wavelength excitation laser, an 1800 lines/mm grating, and 50x objective. High resolution Raman mapping was performed with a 500nm step size using 100x objective. Large area Raman mapping was performed with 633 nm laser, 1800 lines/mm grating, and 50x objective.

ToF-SIMS measurements were conducted with a TOF-SIMS 5 instrument (ION-TOF, Münster, Germany) using a 25 kV Bismuth primary ions. The ToF-SIMS elemental mappings were acquired by operating the instrument in "Burst Alignment" (BA) and "High Current Bunched" (HCBU) mode. During operation the primary ion gun typically scans a field of view of 500×500 µm$^2$ applying a 1024×1024 pixel measurement raster. The BA mode offers a better lateral resolution on the cost of mass resolution. In order to exclude any mass interference the HCBU was used. The charging effects due to $SiO_2$ substrate were compensated by using an electron flood gun with energy of 20 eV. The ToF-SIMS depth profiling was acquired in dual beam mode, by scanning the Bismuth beam over an area of 500×500 µm$^2$ applying a 128×128 pixel measurement raster. An oxygen ion beam with energy of 500 eV was used for material abrasion. This beam scanned an area of 700×700 µm$^2$. These parameters were chosen for best depth resolution. Since the results of such sputtering measurements provide spatially averaged information, only areas with homogeneous Cu coverage in the scanned area (excluding Cu accumulations and "pockets") were selected.

To obtain absolute concentration of metallic residuals and calibrate ToF SIMS results additional measurements were performed using TXRF. TXRF measurements were done on a Bruker AXS TREX 630 tool. A W-K$_\alpha$ x-ray source operated at 40 kV and 40 mA was used at an incident angle of 0.05 °. Analysis was performed both on the areas covered by graphene and uncovered Si surface. To localize the graphene covered areas on the 200 mm Si wafer and guide the TXRF analysis differential work function imaging[35] using a QCept Technologies, Inc. ChemetriQ 5000 was applied (see Supporting Information, Fig S3).

XPS measurements were performed with a PHI VersaProbe II Scanning XPS Microprobe Photoelectron Spectrometer. Long integration times were applied to detect very low intensity metal peaks (if any).

Minority charge carrier diffusion length has been measured using a Semiconductor Diagnostics, Inc. FAaST 230 SPV tool.

**Acknowledgments**

This work was supported by the EU project GRADE 317839, the DFG projects ME 4117/1 and LE 2440/1-1, and the ERC grant INTEGRADE (307311). Partial support by the EU FP7 grant no604391 (Graphene Flagship) is acknowledged. We acknowledge D. Kot of IHP for useful discussions and preparation of reference samples.

# SUPPORTING MATERIAL

# Residual Metallic Contamination of Transferred Chemical Vapor Deposited Graphene

*Grzegorz Lupina,‡\* Julia Kitzmann,‡ Ioan Costina,‡ Mindaugas Lukosius,‡ Christian Wenger,‡ Andre Wolff,‡ Sam Vaziri,† Mikael Östling,† Iwona Pasternak,§ Aleksandra Krajewska,§ Wlodek Strupinski,§ Satender Kataria,⊥ Amit Gahoi,⊥ Max C. Lemme⊥, Guenther Ruhl┬, Guenther Zoth,┬ Oliver Luxenhofer,∥ Wolfgang Mehr‡*

‡IHP, Im Technologiepark 25, 15236 Frankfurt (Oder), Germany

†KTH Royal Institute of Technology, School of ICT, Isafjordsgatan 22, 16440 Kista, Sweden

§Institute of Electronic Materials Technology, Wolczynska 133, 01-919 Warsaw, Poland

⊥University of Siegen, Hölderlinstr. 3, 57076 Siegen, Germany

┬Infineon Technologies AG, Regensburg 93049, Germany

∥Infineon Technologies Dresden GmbH, Dresden 01099, Germany

\*Corresponding Author: Grzegorz Lupina, lupina@ihp-microelectronics.com

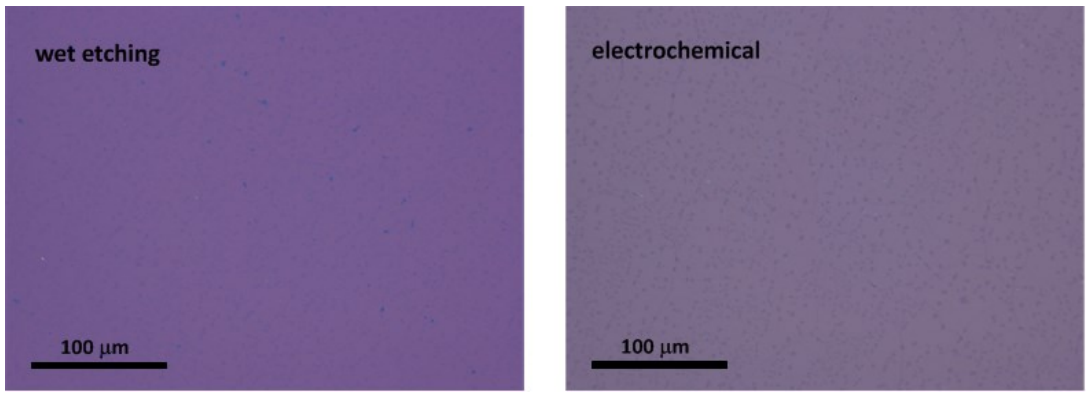

Figure S1. Large area OM images of wet-etched and electrochemically delaminated graphene.

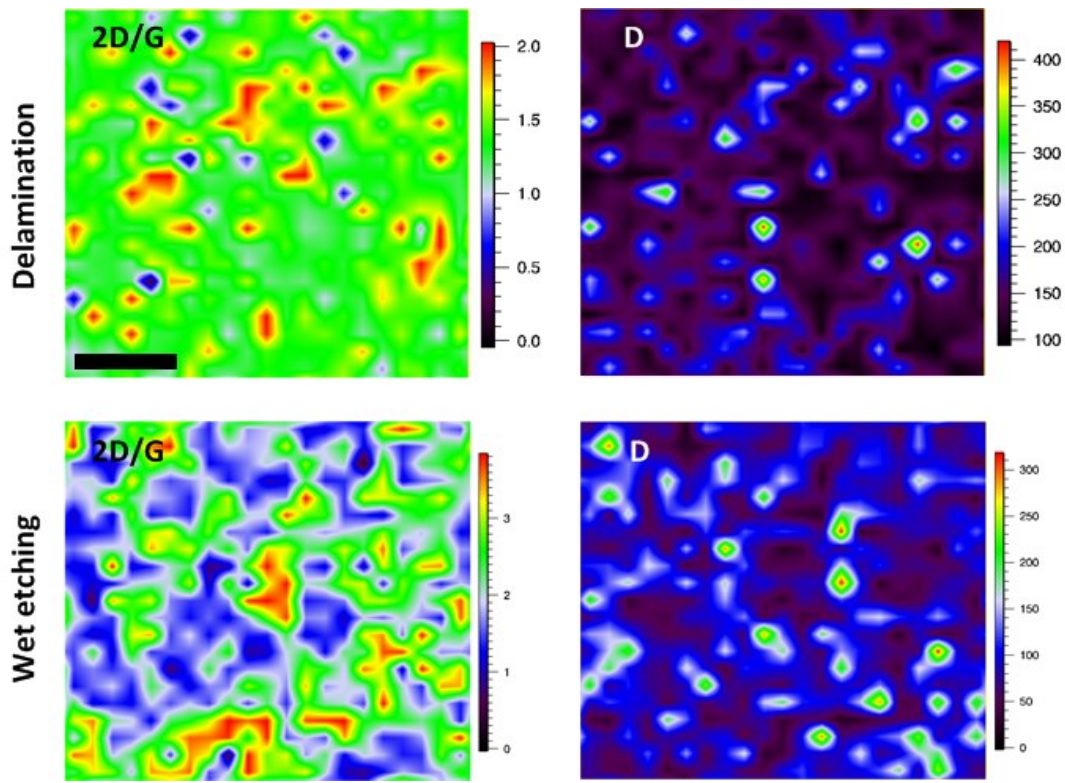

Figure S2. Graphene uniformity: Large area (80x80μm$^2$) Raman mapping on wet-etched and electrochemically delaminated graphene. Scale bar is 20μm.

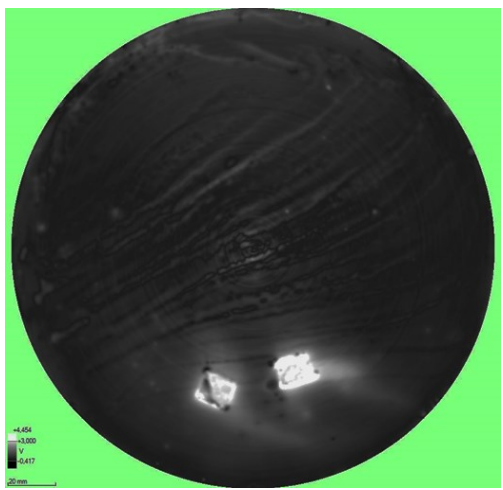

Figure S3. Scanning surface potential difference imaging (differential work function imaging) [Ref. 33] of a 200 mm Si wafer with transferred graphene pieces. The measurements were done using a QCept Technologies, Inc. ChemetriQ 5000 tool. Graphene flakes, which are otherwise poorly optically visible on bare Si(100) surface, are highlighted in surface potential imaging allowing for precise positioning. Acquired position of the flakes was subsequently used to guide TXRF measurements.

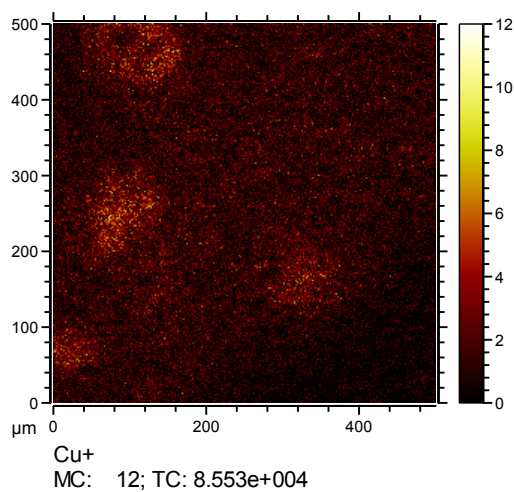

Figure S4. 500x500 $\mu m^2$ ToF SIMS map of Cu+ taken from a graphene sample obtained using electrochemical delamination. Graphene is heavily contaminated with Cu ($\sim 10^{15}$ at./cm2). Accumulations of Cu impurities on a lower intensity background are visible. The origin of such accumulations is not fully understood yet. While redeposition of Cu from etching solution is excluded, features with similar size and distribution were not found on the Cu foil before transfer with EBSD and AFM. Such pockets were not found after subsequent HCl-based cleaning steps anymore.

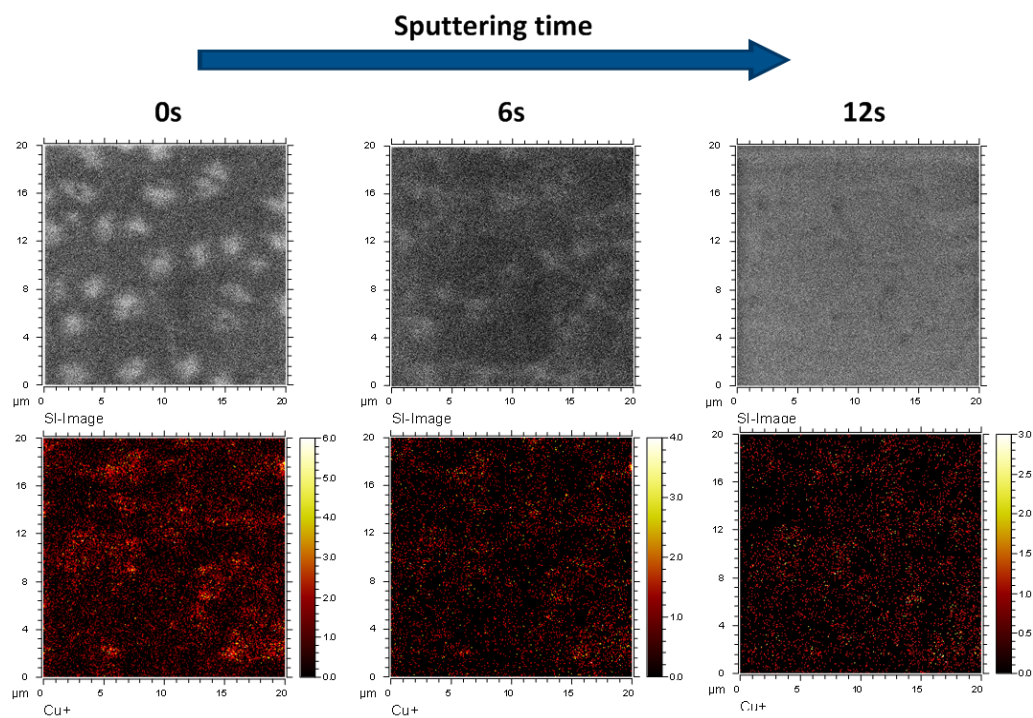

Figure S5. 20x20 μm$^2$ TOF SIMS secondary ion images and Cu distribution maps as a function of sputtering time. Sputtering was performed with a 250eV Cs ion beam.

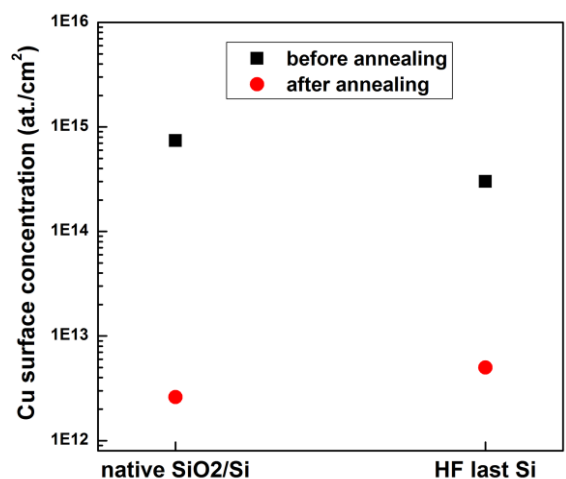

Figure S6. Cu surface concentration as measured by ToF-SIMS for reference samples without graphene. The surface of native SiO$_2$/Si and HF-treated Si wafers was intentionally contaminated using Cu standard solution mixed with DI H$_2$O. Surface concentration of Cu was measured before and after annealing at 500°C for 15 min. For both substrates a reduction of Cu surface concentration by about 2 orders of magnitude was observed.

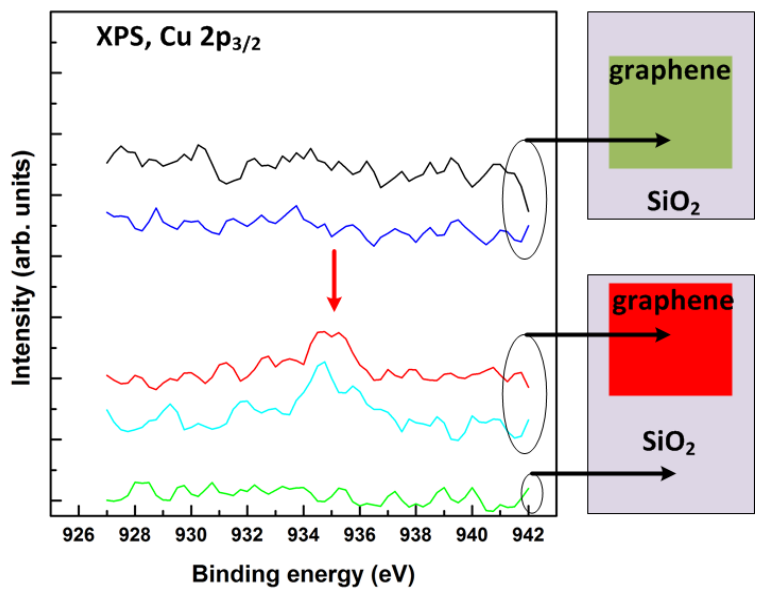

Figure S7. XPS scans in the Cu $2p_{3/2}$ binding energy region for two graphene samples with two different levels of Cu contamination. For heavily contaminated samples (~ $5 \times 10^{14}$ at/cm$^2$) the Cu $2p_{3/2}$ peak is clearly visible on graphene (indicated by arrow). For cleaner graphene samples, the XPS peak associated with Cu is not detected.

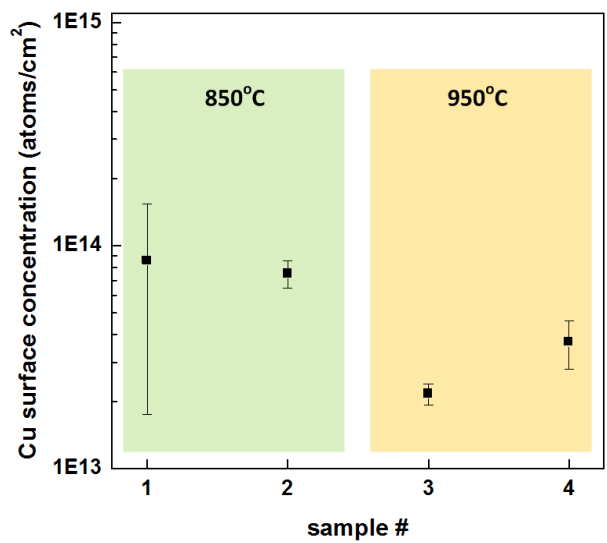

Figure S8. Cu surface concentration measured after transfer to SiO$_2$/Si for CVD graphene grown at 850C and 950C. Graphene was transferred using electrochemical delamination in both cases.

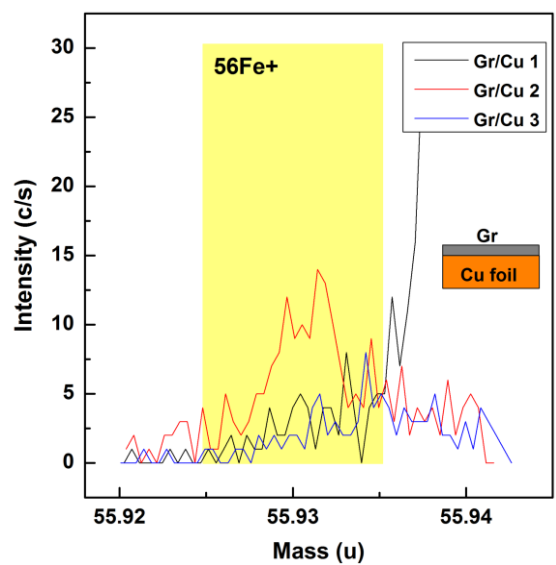

Figure S9. ToF-SIMS mass spectra in the $^{56}$Fe region acquired from the surface of various as-shipped graphene/Cu samples before transfer.